\documentclass[twocolumn,aps,floatfix,superscriptaddress,prl,showpacs]{revtex4}
\usepackage{amsmath, amsthm, amssymb, graphicx,color}
\input epsf.sty
\begin{document}

\newcommand{\BEQ}{\begin{equation}}     
\newcommand{\BEA}{\begin{eqnarray}}
\newcommand{\EEQ}{\end{equation}}       
\newcommand{\EEA}{\end{eqnarray}}
\newcommand{\eps}{\varepsilon}          
\newcommand{\vph}{\varphi}              
\newcommand{\vth}{\vartheta}            
\newcommand{\D}{{\rm d}}
\newcommand{\II}{{\rm i}}               
\newcommand{\arcosh}{{\rm arcosh\,}}    
\newcommand{\erf}{{\rm erf\,}}          
\newcommand{\wit}[1]{\widetilde{#1}}    
\newcommand{\wht}[1]{\widehat{#1}}      
\newcommand{\lap}[1]{\overline{#1}}     
\newcommand{\demi}{\frac{1}{2}}         
\newcommand{\rar}{\rightarrow}          
\newcommand{\gop}{\wht{\phi}_{\vec{0}}} 

\renewcommand{\vec}[1]{{\bf{#1}}}       
\newcommand{\matz}[4] 
     {\mbox{${\begin{array}{cc} #1 & #2 \\ #3 & #4 \end{array}}$}}

\def\g{\gamma}
\def\r{\rho}
\def\w{\omega}
\def\wo{\w_0}
\def\wp{\w_+}
\def\wm{\w_-}
\def\t{\tau}
\def\av#1{\langle#1\rangle}
\def\pf{P_{\rm F}}
\def\pr{P_{\rm R}}
\def\F#1{{\cal F}\left[#1\right]}

\title{Phenomenology of ageing in the Kardar-Parisi-Zhang equation}

\author{Malte Henkel}
\affiliation{Groupe de Physique Statistique, 
D\'epartement de Physique de la Mati\`ere et des Mat\'eriaux, 
Institut Jean Lamour (CNRS UMR 7198), Universit\'e de Lorraine Nancy, 
B.P. 70239, F -- 54506 Vand{\oe}uvre-l\`es-Nancy Cedex, France}

\author{Jae Dong Noh}
\affiliation{Department of Physics, University of Seoul, 
Seoul 130-743, Republic of Korea}
\affiliation{School of Physics, Korea Institute for Advanced 
Study, Seoul 130-722, Republic of Korea}

\author{Michel Pleimling}
\affiliation{Department of Physics, Virginia Tech, Blacksburg, Virginia 24061-0435, USA}

\date{\today}

\begin{abstract}
We study ageing during surface growth processes described by the one-dimensional Kardar-Parisi-Zhang equation.
Starting from a flat initial state, the systems undergo simple ageing in both correlators and linear 
responses and its dynamical scaling is characterised by the
ageing exponents $a=-1/3$, $b=-2/3$, $\lambda_C=\lambda_R=1$ and $z=3/2$.
The form of the autoresponse scaling function is well described
by the recently constructed logarithmic extension of local scale-invariance.
\end{abstract}

   
\maketitle

The study of the motion of interfaces continues as a widely fascinating topic of
statistical physics. One particularly intensively studied case is non-equilibrium growth
processes, which are governed by local rules. Of these, the model equation proposed
by Kardar, Parisi and Zhang (KPZ) \cite{Kard86} continues to play a
paradigmatic role 
in the investigation of the dynamical scaling of such interfaces, 
with an astounding range
of applications, including Burgers turbulence, directed polymers in a random medium, glasses and
vortex lines, domain walls and biophysics, 
see \cite{Bara95,McKa95,Halp95,Krug97,Krie10,Sasa10b,Tong94,Batc00}
for reviews. Remarkably, in $1D$ the height distribution can be shown to converge for large times 
towards the gaussian Tracy-Widom distribution \cite{Sasa10,Cala11}. 
A particularly clean experimental realization of this universality class 
has been found recently in the growing interfaces of turbulent liquid crystals \cite{Take11}.

New insight in the non-equilibrium properties of many-body systems comes from an analysis of the 
{\em ageing properties}, which is realised if the system is rapidly brought out of equilibrium by a change
of one of its state variables \cite{Cugl02,Henk10}. 
By definition, an ageing system (i) undergoes a slow, non-exponential relaxation towards
its stationary state(s), (ii) does not satisfy time-translation-invariance and 
(iii) shows dynamical scaling. Studies of ageing require the analysis of both correlators $C$ and responses $R$ 
to be complete 
and also go beyond the study of dynamics in analysing at least two-time observables. 
Let $s$ denote the waiting time and $t>s$ the observation time. 
For {\em simple ageing}, one expects in the ageing regime $s\gg \tau_{\rm micro}$ and $t-s\gg \tau_{\rm micro}$, 
where $\tau_{\rm micro}$ is a microscopic time scale, a single relevant length scale $L(t)\sim t^{1/z}$ such that
\BEA
C(t,s) &=& \left\langle \phi(t) \phi(s) \right\rangle 
- \left\langle \phi(t)\right\rangle\left\langle \phi(s)\right\rangle 
= s^{-b} f_C\left(\frac{t}{s}\right) 
\label{1} \\
R(t,s) &=& \left. \frac{\delta \left\langle \phi(t)\right\rangle}{\delta
j(s)}\right|_{j=0} 
= \left\langle \phi(t) \wit{\phi}(s) \right\rangle = s^{-1-a} f_R\left(\frac{t}{s}\right) 
\nonumber 
\EEA
where $j$ is the external field conjugate to $\phi$.  
This defines the ageing exponents $a,b$ and the scaling functions, from whose asymptotic behaviour
$f_{C,R}(y) \sim y^{-\lambda_{C,R}/z}$ as $y\to \infty$ one has the autocorrelation and autoresponse exponents
$\lambda_{C,R}$ where $z$ is the dynamical exponent. 
In the context of Janssen-de Dominicis theory, $\wit{\phi}(t)$ is the
response field conjugate to the order-parameter $\phi(t)$. 

For example, simple ageing is found in non-disordered, unfrustrated magnets, quenched from an initial disordered
state to a temperature $T\leq T_c$ at or below its critical temperature $T_c$ 
(see \cite{Henk10} and refs. therein) 
or else in microscopically irreversible systems with a 
non-equilibrium stationary state \cite{Enss04,Rama04,Odor06,Dura11}. 
Generically, one finds $\lambda_C=\lambda_R$, but the values of $a,b$ 
depend more sensitively on the kind of ageing
investigated (for reversible systems on the type of quench and for 
irreversible ones on the specific type of dynamics). 

Here, we shall study what kind of ageing phenomena can arise in the growth of interfaces. 
A typical system is formulated 
in terms of a height variable $h=h_i(t) = h(t,\vec{r}_i)$, 
defined over a substrate in $d$ dimensions. 
A local, microscopic rule
indicates how single particles are added to the surface. 
One of the main quantities studied is the surface roughness 
\BEQ
w^2(t;L) = \frac{1}{L^d} \sum_{i=1}^{L^d} \left\langle \left(h_i(t) - \overline{h}(t) \right)^2 \right\rangle
\EEQ
on a lattice with $L^d$ sites and average height 
{$\overline{h}(t) = L^{-d} \sum_i h_i(t)$}. 
It obeys Family-Vicsek scaling \cite{Fami85}
\BEQ
\hspace{-0.10truecm} w^2(t;L) = L^{2\zeta} f\left(t L^{-z}\right)  , \; 
f(u) \sim \left\{\begin{array}{ll} u^{2\beta} & \mbox{\rm ;\ if $u\ll 1$} \\
                                   \mbox{\rm cste.}  & \mbox{\rm ;\ if $u\gg 1$} 
\end{array} \right.
\EEQ
where $\beta$ is the growth exponent and $\zeta=\beta z$ is the roughness exponent. 
For an infinite system, the width
grows for large times as $w^2(t;\infty) \sim t^{2\beta}$. 

The generic universality class for growth phenomena is given by the KPZ equation \cite{Kard86} 
\BEQ \label{kpz}
\frac{\partial h}{\partial t} = \nu \frac{\partial^2 h}{\partial \vec{r}^2} 
+ \frac{\mu}{2} \left( \frac{\partial h}{\partial \vec{r}}\right)^2 +\eta
\EEQ
where $\eta(t,\vec{r})$ is a white noise with zero mean and variance 
$\langle\eta(t,\vec{r})\eta(t',\vec{r}')\rangle=2\nu T\delta(t-t') \delta(\vec{r}-\vec{r}')$ 
and $\mu,\nu,T$ are material-dependent constants. 
For comparison, we introduce two more universality classes of surface growth: elimination
of the non-linear term in (\ref{kpz}) by setting 
$\mu=0$ 
gives the Edwards-Wilkinson (EW) universality class 
\cite{Edwa82}. The Mullins-Herrings (MH) universality class is given by 
$\partial_t h = - \nu \partial_r^4 h +\eta$ \cite{Wolf90}. For both EW and MH classes, the ageing scaling forms
(\ref{1}) for $C$ and $R$ have been explicitly confirmed \cite{Roet06}. 
Values of some growth and ageing exponents in $1D$ 
are listed in table~\ref{tab1}. For the $1D$ KPZ class, 
the exponents $z,\beta$ are exactly known \cite{Kard86},
whereas the relation $b = - 2 \zeta/z = -2 \beta$ follows from dynamical scaling \cite{Kall99,Daqu11}. Also,
evidence for a growing length $L(t)\sim t^{1/z}$ \cite{Bust07,Chou10} and estimates of $\lambda_C$ \cite{Kall99,Daqu11} 
have been reported~\cite{comment1}. 
However, no systematic test of the ageing scaling has been reported for the space-time correlation function, 
and no information exists at all for the response function $R$. These will be provided now.   

\begin{table}
\begin{tabular}{|c|cccccc|} \hline
model & ~$z$~ & ~$a$~   & ~$b$~  & ~$\lambda_R=\lambda_C$~ & ~$\beta$~ & ~$\zeta$~ \\ \hline
KPZ   & $3/2$ & $-1/3$  & $-2/3$ & $1$                     & $1/3$     & $1/2$       \\
EW    & $2$   & $-1/2$  & $-1/2$ & $1$                     & $1/4$     & $1/2$       \\
MH    & $4$   & $-3/4$  & $-3/4$ & $1$                     & $3/8$     & $3/2$       \\ \hline
\end{tabular}
\caption[Ageing exponents]{Some dynamical, ageing and growth exponents of several universality 
classes in $d=1$ dimension.\label{tab1}}
\end{table}

Our numerical simulations in the $1D$ KPZ class either use the discretised KPZ equation (\ref{kpz}) 
in the strong coupling limit \cite{Newm97} (we checked that our results do not 
depend on the chosen discretisation scheme)
or else the Kim-Kosterlitz (KK) model \cite{Kim89}. This model
uses a height variable $h_i(t)\in\mathbb{Z}$ attached to the sites of a chain with $L$ 
sites and subject to the constraints $|h_i(t)-h_{i\pm 1}(t)|=0,1$, at all sites $i$.
{}From a flat initial condition, that is
$h_i(0)=0$, the dynamics of the model is as follows: at each time step, select randomly a site $i$ and deposit
a particle with probability $p$ or else eliminate a particle with probability $1-p$. 
$L$ such deposition attempts make
up a Monte Carlo step. It is well-known that this model is in the KPZ universality class. 
The choice of the value of $p$ is a practical matter. In order to avoid meta-stable states, we have chosen
$p=0.98$. In simulations, we have taken $L=2^{17}$ and all the 
data have been averaged over $10^4$ samples.
For the discretised KPZ equation we considered systems of size $L = 10^4$ and averaged over
typically $10^5$ samples.

In studying the ageing behaviour, we shall consider the two-time spatio-temporal correlator
\BEA
C(t,s;\vec{r}) &=& 
\left\langle \left( h(t,\vec{r}+\vec{r}_0) - {\left\langle
\overline{h}(t)\right\rangle} \right)\left( h(s,\vec{r}_0) -
{\left\langle \overline{h}(s)\right\rangle} \right) 
\right\rangle 
\nonumber \\
&=& \left\langle h(t,\vec{r}+\vec{r}_0) h(s,\vec{r}_0) \right\rangle -
{\left\langle \overline{h}(t)\right\rangle \left\langle
\overline{h}(s)\right\rangle} 
\nonumber \\
&=& s^{-b} F_C\left( \frac{t}{s}, \frac{|\vec{r}|^z}{s} \right) \label{5}
\EEA
along with the extended Family-Vicsek scaling in the $L\to\infty$ limit and 
where the definition of the exponents
is analogous to the usual one for simple ageing. The autocorrelation exponent can be found from
$f_C(y) = F_C(y,0)\sim y^{-\lambda_C/z}$ as $y\to\infty$. We also have $b=-2\beta$, since the width 
$w^2(t;\infty)=C(t,t;\vec{0})=t^{-b} F_C(1,0)$. This is justified since the initial conditions in the
$1D$ KPZ do not generate new, independent renormalisations \cite{Krec97}.  

\begin{figure}[t] 
\centerline{\epsfxsize=3.20in\ \epsfbox{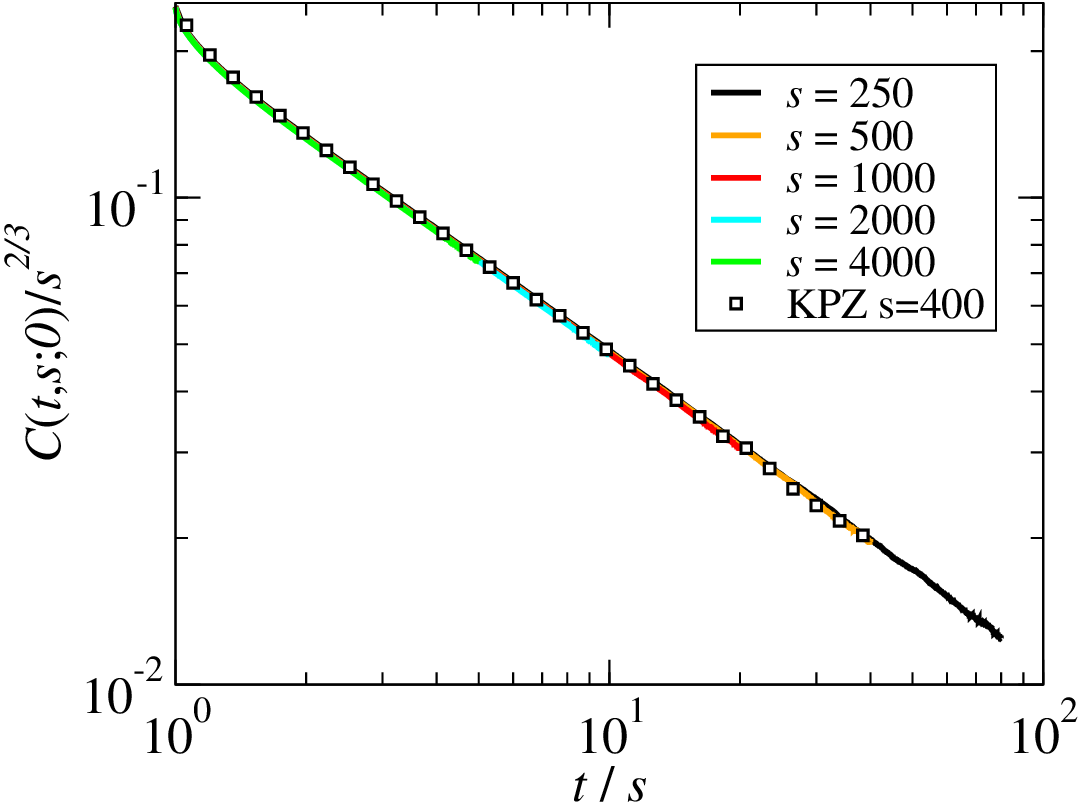}}
\caption{Scaling of the two-time autocorrelator $C(t,s)$ from the KK model and several
values of $s$, and the discretised KPZ equation, with 
$s=400$, rescaled by a factor $2.79$. \label{fig1}}
\end{figure}

In figure~\ref{fig1}, we show data for the autocorrelator $C(t,s)=C(t,s;\vec{0})$ obtained from the KK model. 
A clear data collapse is seen and for large values of the
scaling variable $y=t/s$, an effective power-law behaviour with an exponent 
$\lambda_C/z\approx \frac{2}{3}$ is found. 
The data are fully compatible with a numerical solution of the KPZ equation 
and directly test simple ageing (\ref{1}) in the $1D$ KPZ class.
All this, completely analogous to the EW and MH classes, confirms and 
strengthens earlier conclusions \cite{Kall99,Bust07,Chou10,Daqu11,Krec97}. 

In order to define a response, we appeal to the procedures used in irreversible systems
\cite{Enss04,Odor06,Dura11,Igua09} where the external field is related to a local change of rates. 
In the KK model, we
consider a space-dependent
deposition rate $p_i=p_0 + a_i \eps /2$ with $a_i=\pm 1$ and $\eps=0.005$ a small parameter. 
Then consider, with the {\em same} stochastic noise $\eta$, two
realisations: system A evolves, up to the waiting time $s$, with the site-dependent deposition rate $p_i$ and 
afterwards, with the uniform deposition rate $p_0$. 
System B evolves always with the uniform deposition rate $p_i=p_0$.
Of course, the evaporation rate $q_i=1-p_i$. Then, the time-integrated response function is
\BEA
\lefteqn{ \chi(t,s;\vec{r}) = \int_0^s \!\!\D u\: R(t,u;\vec{r}) }  \label{6} \\
&=& 
\frac{1}{L} \sum_{i=1}^L \left\langle \frac{h_{i+r}^{(A)}(t;s) -
h_{i+r}^{(B)}(t)}{\eps a_i}\right\rangle  
= s^{-a} F_{\chi}\left( \frac{t}{s}, \frac{|\vec{r}|^z}{s}\right)
\nonumber
\EEA
together with the expected scaling. 
The time-integrated autoresponse $\chi(t,s)=\chi(t,s;\vec{0})$ plays the same 
role as the thermoremanent integrated response of 
magnetic systems \cite{Henk10}. The autoresponse exponent is read off from 
$f_{\chi}(y)=F_{\chi}(y,0)\sim y^{-\lambda_R/z}$ for $y\to\infty$. 
For the discretised KPZ equation we realize the perturbation by adding 
a spatially random force, of strength $\pm f_0=\pm 0.3$,
up to the waiting time $s$.

\begin{figure}[tb] 
\centerline{\epsfxsize=3.20in\ \epsfbox{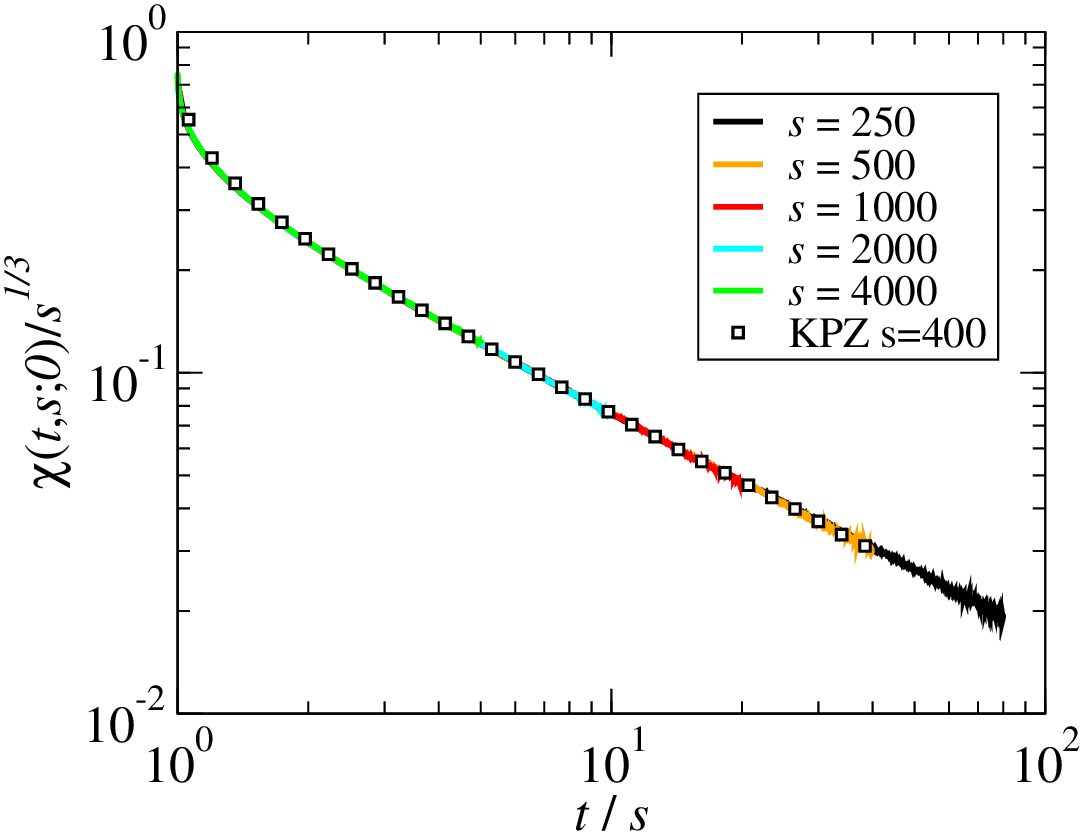}}
\caption{Scaling of the two-time integrated autoresponse $\chi(t,s)$ from the KK model
for several values of $s$, and from the KPZ equation with 
$s=400$, rescaled by a factor $7.25$. \label{fig2}}
\end{figure}

In figure~\ref{fig2}, data for the integrated autoresponse $\chi(t,s)$ 
coming from the KK model are shown. An excellent collapse 
is found for $a=-\frac{1}{3}$. The effective power law, for $y=t/s$ large, 
reproduces well the expected $\lambda_R/z\approx \frac{2}{3}$. 
Indeed, from the exact fluctuation-dissipation theorem $TR(t,s;r)=-\partial_r^2 C(t,s;r)$, 
valid in the $1D$ KPZ universality class (because of time-reversal-invariance) 
\cite{Deke75,Lvov93,Frey96,Cane10}, 
we obtain the predictions $1+a=b+2/z$ and $\lambda_C=\lambda_R$, in agreement with our data.
The data are essentially identical to those
obtained from the KPZ equation, in agreement with universality. For the first time, the ageing form (\ref{1}) of
the linear response is confirmed in a non-linear growth model. In contrast with the EW and MH classes,
$a$ and $b$ are different, a feature commonly seen in irreversible systems \cite{Henk07,Henk10}.  

\begin{figure}[tb] 
\centerline{\epsfxsize=3.20in\ \epsfbox{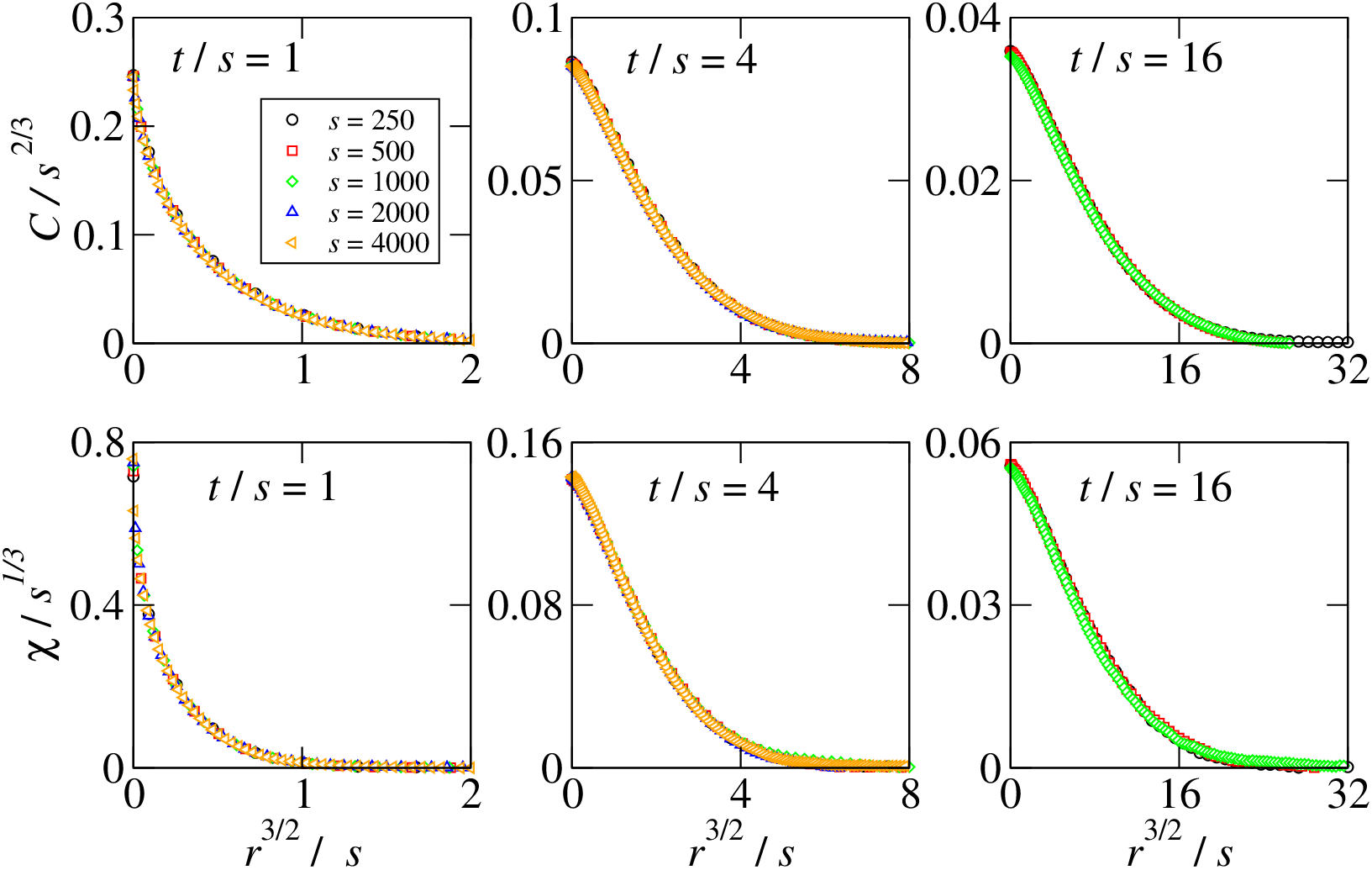}}
\caption{Space-dependent scaling of the two-time correlator $C(t,s)$ (upper row) and the integrated response
$\chi(t,s)$ (lower row), for several values of $s$ and the scaling variable $y=t/s$. \label{fig3}}
\end{figure}

Next, in figure~\ref{fig3}, we illustrate the space-dependent scaling 
of both the correlator and the integrated response. 
For several values of the scaling variable $y=t/s$, the dependence on the second argument in the scaling forms
(\ref{5},\ref{6}) is illustrated. An excellent data collapse is found, 
which further confirms the conclusions already
drawn from the autocorrelator and the autoresponse and also confirms the 
exactly known dynamical exponent $z=\frac{3}{2}$
in the $1D$ KPZ universality class. The shape of the scaling functions changes notably when $y$ is varied. 

We now turn to an analysis of the {\em form} of the autoresponse scaling function $f_{\chi}(y)$. 
For ageing simple magnets 
(i.e. non-disordered and unfrustrated), 
it has been proposed to generalise dynamical scaling to a larger set 
of local scale-transformations \cite{Henk94},
which includes the transformation $t\mapsto t/(1+\gamma t)$. 
This hypothesis of local scale-invariance (LSI) indeed reproduces 
precisely the universal shapes of responses and 
correlators in a large variety of models, as reviewed in detail in \cite{Henk10}. Analogous
evidence exists in some irreversible models \cite{Enss04,Odor06,Henk07,Henk10}.  
Similarly, the responses and correlators in the EW and MH classes, with the the local height variable 
$h(t,\vec{r})-\overline{h}(t)$ acting as a quasi-primary scaling operator, 
are described by LSI \cite{Roet06,Henk07b}. {\it Is LSI also realised in the $1D$ KPZ class, 
with the local height as a quasi-primary operator~?}  

We concentrate on the autoresponse function and shall restrict attention to the
transformations in time. The transformation $\delta\phi = \eps X_n \phi$ of the quasi-primary operators 
under local scale-transformations is given by the infinitesimal generators $X_{n}$, which read \cite{Henk06}
\BEQ
X_n = - t^{n+1}\partial_t - (n+1) \frac{x}{z} t^n - n \frac{2\xi}{z} t^n \;\; , \;\; n\geq 0
\EEQ 
and satisfy the commutator $[X_n,X_m]=(n-m)X_{n+m}$. We merely look 
at the finite-dimensional sub-algebra spanned
by the dilatations $X_0$ and the special transformations $X_1$. 
Since time-translations (generated by $X_{-1}$) are absent, we have two {\em distinct}
scaling dimensions $x$ and $\xi$, which together give the shape of the autoresponse function, see below. 
Now, consider a possible extension to so-called logarithmic form, 
where a primary operator $\phi$ is replaced
by a doublet {\small $\left(\begin{array}{c} \phi \\ \psi\end{array}\right)$}. 
In analogy with logarithmic conformal invariance~\cite{Gura93,comment2}, 
this extension is formally carried out by replacing the scaling dimensions $x,\xi$ by matrices  
(restricted to the $2\times 2$ case) \cite{Henk10b}
\BEQ 
x \mapsto \left(\matz{x}{x'}{0}{x}\right) \;\; , \;\;
\xi  \mapsto \left(\matz{\xi}{\xi'}{\xi''}{\xi}\right)
\EEQ
where the first scaling dimension is immediately taken in a Jordan form. 
Consistency with the commutators then leads to $\xi''=0$ \cite{Henk10b}. Recalling (\ref{1}), 
consider the fol\-low\-ing quasi-primary two-point functions, with $y=t/s$
\BEA
\left\langle \phi(t)\wit{\phi}(s)\right\rangle &=& 
s^{-(x+\wit{x})/z}\: {\cal F}(y)  f_0 \nonumber \\
\left\langle \phi(t)\wit{\psi}(s)\right\rangle &=& 
s^{-(x+\wit{x})/z}\: {\cal F}(y)
\Bigl( g_{12}(y) + \gamma_{12} \ln s \Bigr) 
\nonumber \\
\left\langle \psi(t)\wit{\phi}(s)\right\rangle &=& 
s^{-(x+\wit{x})/z}\: {\cal F}(y)
\Bigl( g_{21}(y) + \gamma_{21} \ln s  \Bigr)   
\nonumber \\
\left\langle \psi(t)\wit{\psi}(s)\right\rangle &=& 
s^{-(x+\wit{x})/z} \: {\cal F}(y) \sum_{j=0}^2 h_j(y) \ln^j s 
\EEA
where ${\cal F}(y)=y^{(2\wit{\xi} +\wit{x}-x)/z} (y-1)^{-(x+\wit{x}+2\xi+2\wit{\xi})/z}$ 
and explicitly known scaling functions \cite{Henk10b}. 
In contrast to logarithmic conformal invariance, 
logarithmic corrections to scaling are absent if $x'=\wit{x}'=0$ and there
are no logarithmic factors for $y\to\infty$ if furthermore $\xi'=0$. If we take 
$R(t,s) = \left\langle \psi(t)\wit{\psi}(s)\right\rangle=s^{-1-a} f_R(t/s)$, we find 
\BEA
\lefteqn{ f_R(y) = y^{-\lambda_R/z} \left( 1 - y^{-1}\right)^{-1-a'} } \nonumber \\
&\times& \left[ h_0 - g_0 \ln\left( 1 - y^{-1}\right) - \demi f_0  \ln^2\left( 1 - y^{-1}\right)\right] 
\label{10}
\EEA
with the exponents $1+a = (x + \wit{x})/z$, $a'-a = \frac{2}{z} \left( \xi + \wit{\xi}\,\right)$, 
$\lambda_R/z = x + \xi$ and the normalisation constants $h_0, g_0,f_0$. 

The integrated autoresponse $\chi(t,s) = s^{-a} f_{\chi}(t/s)$ 
is found from (\ref{10}) by using the 
specific value $\lambda_R/z-a=1$ which holds true for the $1D$ KPZ. We find 
\BEA 
\lefteqn{ f_{\chi}(y) = y^{+1/3} 
\left\{ A_0 \left[ 1 - \left( 1- y^{-1}\right)^{-a'} \right] \right. } \label{12} \\
&+& \left.  \left( 1 - y^{-1}\right)^{-a'} \left[ A_1 \ln\left( 1 - y^{-1}\right) 
+ A_2  \ln^2\left( 1 - y^{-1}\right) \right] \right\}
\nonumber
\EEA
where $A_{0,1,2}$ are normalisations related to $f_0,g_0,h_0$.
The non-logarithmic case is recovered if $A_1=A_2=0$. Indeed, for $y\gg 1$, 
one has $f_{\chi}(y) \sim y^{-2/3}$, as it should be.

\begin{figure}[tb] 
\centerline{\epsfxsize=3.20in\ \epsfbox{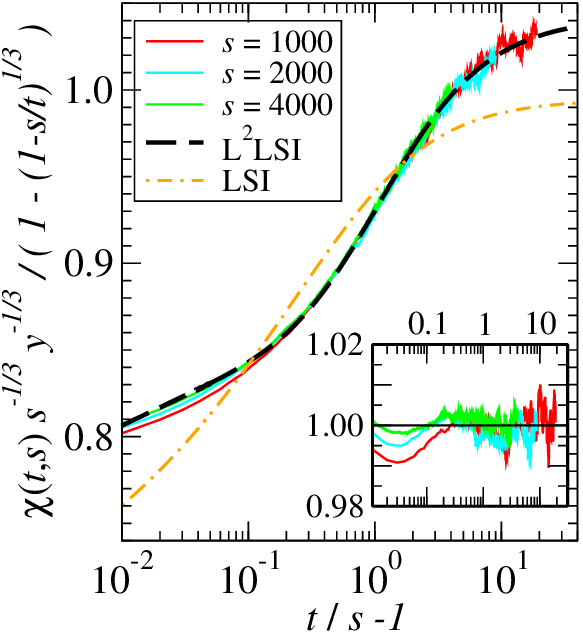}}
\caption{Comparison of the reduced scaling function 
$f_{\rm red}(y) = f_{\chi}(y) y^{-1/3} (1-(1-1/y)^{1/3})^{-1}$ 
of $\chi(t,s)=s^{1/3}f_{\chi}(t/s)$ 
with logarithmic local scale-invariance. Non-logarithmic LSI gives the dash-dotted curve labelled LSI and
full logarithmic LSI (\ref{12}) gives the dashed curve labelled L$^2$LSI.
The inset shows the ratio $\chi(t,s)/\chi_{\mbox{\rm\footnotesize L$^2$LSI}}(t,s)$ over against $t/s-1$.
\label{fig4}}
\end{figure}

In figure~\ref{fig4}, which gives a more fine appreciation of the shape of $f_{\chi}(y)$ than figure~\ref{fig2}, 
we compare data for the reduced scaling function 
$f_{\rm red}(y) = f_{\chi}(y) y^{-1/3} \left[1-(1-y^{-1})^{1/3}\right]^{-1}$ 
with the predicted form (\ref{12}). Data with $s<10^3$ are not yet fully in the scaling regime. 
If one tries to fit the data with a non-logarithmic LSI (then $R=\langle \phi\wit{\phi}\rangle$ or 
$\langle \psi\wit{\phi}\rangle$) one obtains an agreement with the data, 
with a numerical precision of about 5\%. 
An attempt to fit only with the first-order logarithmic terms 
(then $R=\langle \phi\wit{\psi}\rangle$) with $A_2=0$ assumed,
gives back the same result, see table~\ref{tab2}. Only if one uses the full structure of logarithmic LSI,
an excellent representation of the data is found, 
to an accuracy better than $0.1\%$ over the range of data available. 
In the inset the ratio $\chi(t,s)/\chi_{\mbox{\rm\footnotesize L$^2$LSI}}(t,s)$ is shown and we see that at least down
to $t/s\approx 1.03$, the data collapse indicating dynamical scaling holds true, within  the accuracy limits set by the
stochastic noise, within $\approx 0.5\%$. For the largest waiting time $s=4000$, this observation entends over the
entire range of values of $t/s$ considered. 
This indicates that the local height $h$ and its response field $\wit{h}$ of the $1D$ KPZ equation 
could be tentatively identified with the 
logarithmic quasi-primary operators $\psi$, $\wit{\psi}$, 
which slightly generalises the findings for the EW and MH classes,
which obey non-logarithmic LSI. 
It is an open question whether
the approach outlined here just generates the first two terms of an infinite
logarithmic series in $R(t,s)$.

A systematic analysis of the invariance properties of the dynamical functionals studied for instance in \cite{Krec97,Lvov93,Frey96},
or the alternate form derived in \cite{Wio09}, would be of interest, following the lines of study for the analysis of dynamical
symmetries in phase-ordering kinetics, see \cite{Henk10} and refs. therein.

\begin{table}
\begin{tabular}{|l|llll|} \hline
                 & \multicolumn{4}{c|}{parameters} \\ \hline 
$R$                                         & ~~$a'$    & ~$A_0$   & ~~~$A_1$          & ~~~$A_2$   \\ \hline 
$\langle\phi\wit{\phi}\rangle$  -- LSI      & $-0.500$  & $0.662$  & ~~0               & ~~0        \\
$\langle\phi\wit{\psi}\rangle$  -- L$^1$LSI & $-0.500$  & $0.663$  & $-6\cdot 10^{-4}$ & ~~0        \\
$\langle\psi\wit{\psi}\rangle$  -- L$^2$LSI & $-0.8206$ & $0.7187$ & $~~0.2424$        & $-0.09087$ \\ \hline
\end{tabular}
\caption[Fit parameters]{Fitted parameters $A_{0,1,2}$ and $a'$ used in figure~\ref{fig4}. \label{tab2}} 
\end{table}

Summarising, we tested the full scaling behaviour of simple ageing, 
both for correlators and responses, of systems in the $1D$ KPZ universality class. 
This is the first example of a
growth process described by a non-linear equation which is shown to satisfy simple ageing
scaling for space- and time-dependent quantities. 
It is non-trivial that the values of the growth and dynamical exponents, previously known from the study of
the {\em stationary state}, are confirmed {\em far} from stationarity.
It would be interesting to measure it also experimentally. 
Performing a numerical experiment, we find the form of the autoresponse scaling function 
to be very well described 
by the recently constructed logarithmic extension of local scale-invariance, with
a natural identification of the leading quasi-primary operators.  
In view of important recent progress in the exact solution of the $1D$ KPZ equation, see \cite{Sasa10b,Sasa10,Cala11}, 
one may expect that the question of
a logarithmic dynamical scaling can be adressed and its further consequences explored.

This work was supported by the US National
Science Foundation through DMR-0904999. 
This work was also supported by Mid-career Researcher Program
through NRF grant No.~2011-0017982 funded by the Ministry of Education,
Science, and Technology of Korea.


\begin{thebibliography}{99}
\bibitem{comment1} In the EW and MH universality classes, one has $\lambda_C=\lambda_R=d$. 
For the KPZ class $\lambda_C=d$ is conjectured \cite{Kall99}.
\end{thebibliography}
\end{document}